\documentclass{PoS}
\usepackage{epsfig}
\usepackage{graphicx}
\usepackage{dcolumn}
\usepackage{bm}

\newcommand{\be}{\begin{equation}}
\newcommand{\ee}{\end{equation}}
\newcommand{\ba}{\begin{eqnarray}}
\newcommand{\ea}{\end{eqnarray}}
\newcommand{\bi}{\begin{itemize}}
\newcommand{\ei}{\end{itemize}}
\newcommand{\tr}{{\rm Tr\,}}
\newcommand{\re}{\mathop{\rm Re}}

\newcommand{\ovl}{\overline}

\newcommand{\half}{{\textstyle\frac{1}{2}}}

\newcommand{\quarter}{{\textstyle\frac{1}{4}}}

\newcommand{\<}{\langle}
\renewcommand{\>}{\rangle}
\newcommand{\eq}{Eq.~}

\newcommand{\tab}{Tab.~}
\newcommand{\la}{\label}

\newcommand{\txts}{\textstyle}

\title{The Glue Content of the Pion}

\ShortTitle{Glue Content of the Pion}

\author{\speaker{Harvey B. Meyer},~ John W. Negele\\ 
      Center for Theoretical Physics\\
      Massachusetts Institute of Technology \\
      Cambridge, MA 02139 U.S.A.\\
      E-mail: \email{meyerh@mit.edu, negele@mit.edu}}

\abstract{We perform a quenched computation of the glue momentum fraction in the pion. 
Different discretizations of the gluonic energy-momentum tensor 
are studied on the lattice for that purpose. 
We discuss some implications based on the momentum sum rule. 
Finally we point out promising applications of the techniques developed here.

\vspace{0.4cm}

MIT-CTP 3872}

\FullConference{The XXV International Symposium on Lattice Field Theory\\
		 July 30-4 August 2007\\
		 Regensburg, Germany}

\begin{document}

\section{Introduction}
A striking feature of QCD is the large contribution of gluons to the mass and
momentum of hadrons, so it is of fundamental interest to calculate the contributions of
gluons from first principles using lattice QCD. The first moments
\ba
\<x\>_{\rm f}(q^2) &\equiv& {\txts\sum_{f=u,d,s}\int_0^1} xdx \left\{\bar f(x,q^2) + f(x,q^2)\right\} \\
\<x\>_{\rm g}(q^2) &\equiv& {\txts \int_0^1} xdx ~ g(x,q^2)
\ea
of the quark and gluon distribution functions $f(x),~\bar f(x)$ ($f=u,d,s,\dots$) and $g(x)$
acquire a precise field-theoretic meaning via the
operator product expansion in QCD. They
satisfy the well-known momentum sum rule (MSR) $\<x\>_{\rm f}(q^2) +\<x\>_{\rm g}(q^2)=1$
and are related to the corresponding contributions to the
energy-momentum tensor $T_{\mu\nu}$ evaluated on the hadronic state.
Separating the traceless part $\ovl T_{\mu\nu}$ from the trace part $S$
for gluons, denoted `g', and quarks, denoted `f',
$T_{\mu\nu}$ has the explicit form
\ba
T_{\mu\nu} \! & \equiv & \!  \overline T_{\mu\nu}^{\rm g} +
                  \overline T_{\mu\nu}^{\rm f}
               + \quarter\delta_{\mu\nu}(S^{\rm g} + S^{\rm f}), \\
\overline T_{\mu\nu}^{\rm g}\! &=& \!
{\txts\frac{1}{4}}\delta_{\mu\nu}F_{\rho\sigma}^a F_{\rho\sigma}^a
   - F_{\mu\alpha}^a F_{\nu\alpha}^a ,\\
\overline T_{\mu\nu}^{\rm f} \!&=& \!
\quarter {\txts\sum_f} \bar\psi_f\!\! \stackrel{\leftrightarrow}{D_{\mu}}\!\gamma_{\nu}\psi_f
+ \bar\psi_f\!\! \stackrel{\leftrightarrow}{D_{\nu}}\!\gamma_{\mu}\psi_f
 -{\txts\frac{1}{2}}\delta_{\mu\nu}  \bar\psi_f
\! \stackrel{\leftrightarrow}{D_{\rho}}\! \gamma_{\rho}\psi_f ,\\
S^{\rm g} &=& \beta(g)/(2g) ~ F_{\rho\sigma}^a  F_{\rho\sigma}^a,\quad
S^{\rm f} =  [1+\gamma_m(g)] {\txts \sum_f} \bar\psi_f m\psi_f
\ea
where $\stackrel{\leftrightarrow}{D_{\mu}}=
\stackrel{\rightarrow}{D_{\mu}} - \stackrel{\leftarrow}{D_{\mu}} $,
$\beta(g)=-b_0g^3+\dots$  is the beta-function with 
$b_0=(\frac{11}{3}N-\frac{2}{3}N_{\rm f})(4\pi)^{-2}$, 
$\gamma_m(g)$  is the anomalous dimension
of the mass operator,  and all expressions are written in Euclidean space.
The gauge action in this notation is $\frac{1}{4}F_{\mu\nu}^aF_{\mu\nu}^a$.
For an on-shell particle with four-momentum $p=(iE_p,{\bf p})$,
$E^2_{\bf p} = M^2 + {\bf p}^2$, we have the relations
\ba
 \< \Psi,{\bf p}|{\txts\int}\! d^3{\bf z}\,\overline T_{00}^{\rm f,g}(z)\, | \Psi,{\bf p}\>
\!&=&\! [E_{\bf p} -\quarter M^2/E_{\bf p}]~ \<x\>_{\rm f,g},\quad \la{eq:x}\\
 \< \Psi,{\bf p}|{\txts\int}\! d^3{\bf z}\,S^{\rm f,g}(z)\, | \Psi,{\bf p}\>
\!&=&\! (M^2/E_{\bf p}) ~b_{\rm f,g}, \la{eq:b}\\
\<x\>_{\rm f} + \<x\>_{\rm g} &=& b_{\rm f} + b_{\rm g} = 1,  \la{eq:xb}
\ea
where states are normalized according to $\<{\bf p}|{\bf p} \> = 1 $.
The renormalization of $\<x\>_{\rm f,g}$ will be discussed in section~\ref{sec:renorm}.

Equation~\ref{eq:x}  shows that in the infinite momentum frame, where $E_p \sim P \to \infty$,
$\langle x \rangle_g$ represents the momentum fraction arising from  gluons,
and calculating $\langle x \rangle_g$ is the main goal of this work.
In the rest frame, the gluon contribution of Eq.~\ref{eq:x} to the hadron mass
is $\frac{3}{4} M   \langle x \rangle_g$~\cite{ji}.   From Eq.~\ref{eq:b} in the rest frame,
the contribution of the trace anomaly $S^g$ to the hadron mass is
$\frac{1}{4}b_g M$~\cite{ji}, and we have  performed
the first step to calculate this matrix element as well~\cite{gluex}.
Finally, we remark that in thermodynamics at temperature $T$ 
the energy density $\epsilon$ and pressure $P$ are given by 
\be
\epsilon-3P = \<\,S\,\>_T -\<\,S\,\>_0,
  \qquad\qquad
\epsilon+P = {\txts\frac{4}{3}}\<\,\ovl T_{00}\,\>_T  ~.\la{eq:basic}
\ee

The pioneering calculation of the glue momentum fraction 
(in the proton) was presented at the LATTICE96 conference~\cite{horsley}.
In the present study we treat the case of ``heavy pions'' with masses in the range
$600\, {\rm MeV}<M_\pi<1060 \,{\rm MeV}$. Our improved techniques,
applied here in the quenched approximation, are applicable
in full QCD calculations, and to the case of the proton.

\section{Discretization and normalization of the gluonic operators}

 We use the Wilson gluon action
$ \frac{1}{g_0^2} \sum_{x,\mu\neq\nu} \tr\{1-P_{\mu\nu}(x)\}$
 and the Wilson fermion action~\cite{wilson74}.
There are two distinct ways~\cite{liu} to discretize the Euclidean gluonic operator
$\overline T^{\rm g}_{00} = \half (-{\bf E}^a \cdot {\bf E}^a + {\bf B}^a \cdot {\bf B}^a)$
and the trace anomaly
\mbox{$S^{\rm g} = \frac{\beta(g)}{g} ({\bf E}^a \cdot {\bf E}^a + {\bf B}^a \cdot {\bf B}^a)$ }
on a hypercubic lattice.

The first, denoted `bp' for bare-plaquette,  uses a sum of bare plaquettes
$P_{\mu\nu}$ around a body-centered point $x_\odot = x + \half a \sum_\mu \hat\mu$:

\ba  \la{eq:T00plaq} 
\overline T_{00}^{\rm bp}(x_\odot)
&\equiv &\chi^{\rm bp}(g_0) \frac{Z_{\rm g}(g_0)}{8a^4g_0^2}
\sum_{\cal P} {s}_{00}(0_{\cal P},1_{\cal P})
\sum_{\omega,\lambda=0,1}
\tr\{1- P_{0_{\cal P} 1_{\cal P}}
(x+a\lambda\widehat{2_{\cal P}} +a\omega\widehat{3_{\cal P}})\}
 \\
     \la{eq:Splaq}
S^{\rm bp}(x_\odot)
&\equiv &\chi_s^{\rm bp}(g_0) \frac{dg_0^{-2}}{d\log a}  \frac{1}{8a^4}
\sum_{\cal P}
\sum_{\omega,\lambda=0,1}
\tr \{ 1- P_{0_{\cal P} 1_{\cal P}}
(x+a\lambda\widehat{2_{\cal P}} +a\omega\widehat{3_{\cal P}})\}
\ea
where  $\mu_{\cal P}$ is the image under permutation ${\cal P}$ of $\mu$ and
${s}_{00}(\mu,\nu)$ is 1 if $(\mu,\nu)$ are both spatial and -1 otherwise.
Other diagonal elements of the energy-momentum tensor
are obtained by letting the other directions play the role of time.
When summed over a time-slice, these expressions simplify to Eq. 7 of Ref.~\cite{gluex}.
The other form (`bare clover') is
\ba   \la{eq:T00clover}
\overline T_{00}^{\rm bc}(x)
&\equiv& \frac{\chi^{\rm bc}(g_0) Z_{\rm g}(g_0)}{g_0^2}
\re\tr\Big[\sum_{k} (\widehat F_{0k})^2
      -  \sum_{k<l}(\widehat F_{kl})^2   \Big] \\
\la{eq:Sclover}
S^{\rm bc}(x)&\equiv & \chi^{\rm bc}_s(g_0) \frac{dg_0^{-2}}{d\log a}
\re\tr\Big[\sum_{k} (\widehat F_{0k})^2
      +  \sum_{k<l}(\widehat F_{kl})^2   \Big],
\ea
where $\widehat F_{\mu\nu}(x)$ is the clover-shaped discretization of
the field-strength tensor (see~\cite{sommer96}).
This form allows for the discretizations
of off-diagonal elements of $\overline T_{\mu\nu}$ as well.
Each of the normalization factors $Z_{\rm g}(g_0)$, $\chi^{\rm bc}(g_0)$
and $\chi^{\rm bc}_s(g_0)$ in \eq (\ref{eq:T00plaq},\ref{eq:T00clover}) 
is of the form $1+{\rm O}(g_0^2)$. 
The  factor $Z_{\rm g}(g_0)$ 
is dictated by an exact lattice sum-rule for the Wilson gauge action and
is known with a precision of about $1\%$ (see~\cite{hm-visco} and Refs. therein).
To obtain the absolute normalization of other discretizations, it is sufficient
to compute their normalization $ \chi(g_0)$ relative to that of the bare plaquette.

New versions of the gluonic operators are obtained 
by replacing the link variables in
\eq(\ref{eq:T00plaq}--\ref{eq:Sclover}) by  `smeared' versions thereof.
The HYP form of smearing~\cite{hyp} is particularly
attractive in the present context
in that it preserves the symmetry among all four Euclidean directions.

Our criteria for the choice of the discretization are
to maximize the signal-to-noise ratio, minimize cutoff effects,
and preserve locality as much as possible.
We studied the signal-to-noise ratio for four different discretizations
by comparing the variance of the entropy at
temperature $T=1/L_0=1.21T_c$~\cite{teper-sun}, which is proportional to the
expectation value of   $\sum_x \ovl T_{00}(x)$, on  $L_0\times L^3 $ lattices 
of fixed physical size. 

The resulting variances  for the plaquette and clover discretizations
with bare and HYP links are shown in Table \ref{tab:varT00}.
We find  dramatic differences between the discretizations, with
the HYP-clover operator reducing the variance by almost two orders of magnitude
as compared to the bare plaquette operator.
Variance reduction comes at the cost of a certain loss of locality,
since the HYP plaquette and HYP-clover operators have extent $3a$ and $4a$ respectively.
As Fig.~(\ref{fig:variance}, left panel) illustrates, the variance 
of $\sum_x \ovl T^{\rm g}_{00}(x)$ grows like $a^{-4}$, but the prefactor
is non-universal and depends strongly on the discretization.
As a compromise  between locality and variance reduction, from now on
we work with the HYP-plaquette operator.
In~\cite{gluex}, we performed a check of its discretization errors by computing
the dependence of $\chi$ on $a/L_0$ at $\beta=6$.
We showed  that the dependence of $\chi$ on $a/L_0$ 
is mild and statistically consistent with zero for $L_0/a\geq6$.

\begin{table}
\begin{center}
\begin{tabular}{|c|c|c@{~~~}c|c@{~~~}c|}
\cline{3-6}
\multicolumn{2}{c|}{}& \multicolumn{2}{c |}{relative variance}  &  \multicolumn{2}{c|}{normalization} \\
\cline{3-6}
\multicolumn{2}{c|}{}    & bare  &  HYP & bare  &  HYP \\
\hline
$\beta=6.000$& plaq.   & 26.42(71)    & 0.6518(43) &   1.000  & 0.5489(68) \\
 $ 6\times16^3$ & clover  & 3.85(11)    & 0.3049(41) &  2.184(67)  & 0.613(20)  \\
\hline
$\beta=6.093$& plaq.   & 45.5(2.5)    & 0.991(11) &   1.000  & 0.546(14) \\
 $ 7\times20^3$    & clover  & 6.20(14)    & 0.4393(40) &  2.026(52)  & 0.585(16)  \\
\hline
$\beta=6.180$& plaq.   & 81.5(5.8)    & 1.940(26) &   1.000  & 0.596(20) \\
$ 8\times22^3$   & clover  & 11.84(33)    & 0.8280(85) &  2.092(68)  & 0.636(22)  \\
\hline
$\beta=6.295$& plaq.   & 157(16)    & 3.080(50) &   1.000  & 0.563(28) \\
$ 9\times24^3$   & clover  & 19.74(75)    & 1.268(16) &  1.928(93)  & 0.598(30)  \\
\hline
$\beta=6.408$& plaq.   & 346(81)    & 7.67(29) &   1.000  & 0.617(68) \\
$ 11\times29^3$   & clover  & 43.3(3.6)    & 3.042(78) &  1.91(20)  & 0.654(75)  \\
\hline
\end{tabular}
\end{center}
\caption{
\underline{Left}: the relative variance, $\<{\cal O}^2\>/\<{\cal O}\>^2-1$,
of the operators ${\cal O}=\sum_x  \ovl T^{\rm g}_{00}(x)$ on lattices of matched physical sizes
for different discretizations  described in the text.
\underline{Right}:
the normalization $\chi(g_0,a/L_0)$ of the operator relative to the bare plaquette, 
determined on the same lattice.}
\la{tab:varT00}
\end{table}
\begin{figure}
\begin{center}
\begin{minipage}{7cm}
\psfig{file=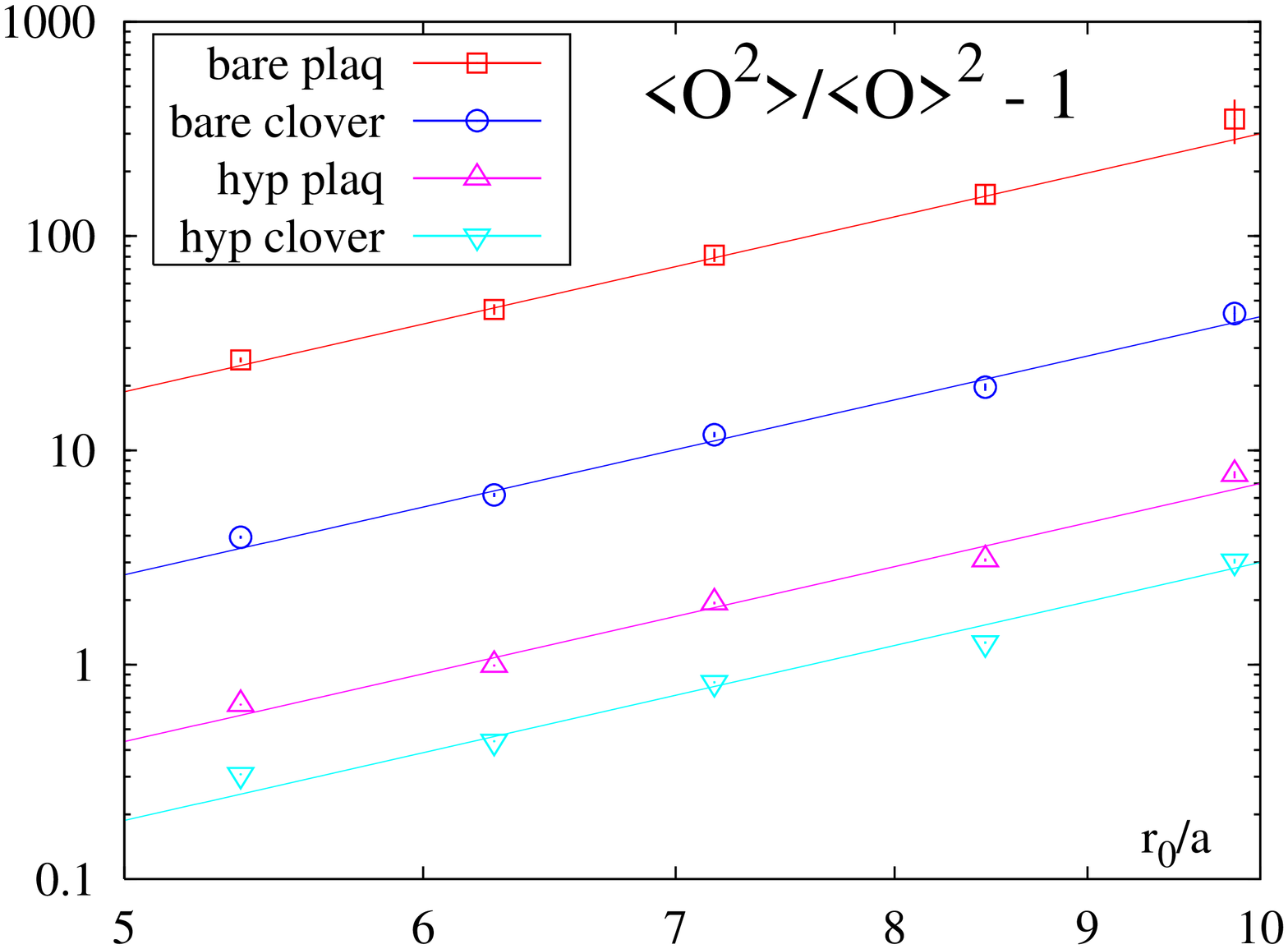,angle=-90,width=7.3cm}
\end{minipage}
\begin{minipage}{8cm}
\vspace{-0.5cm}
\psfig{file=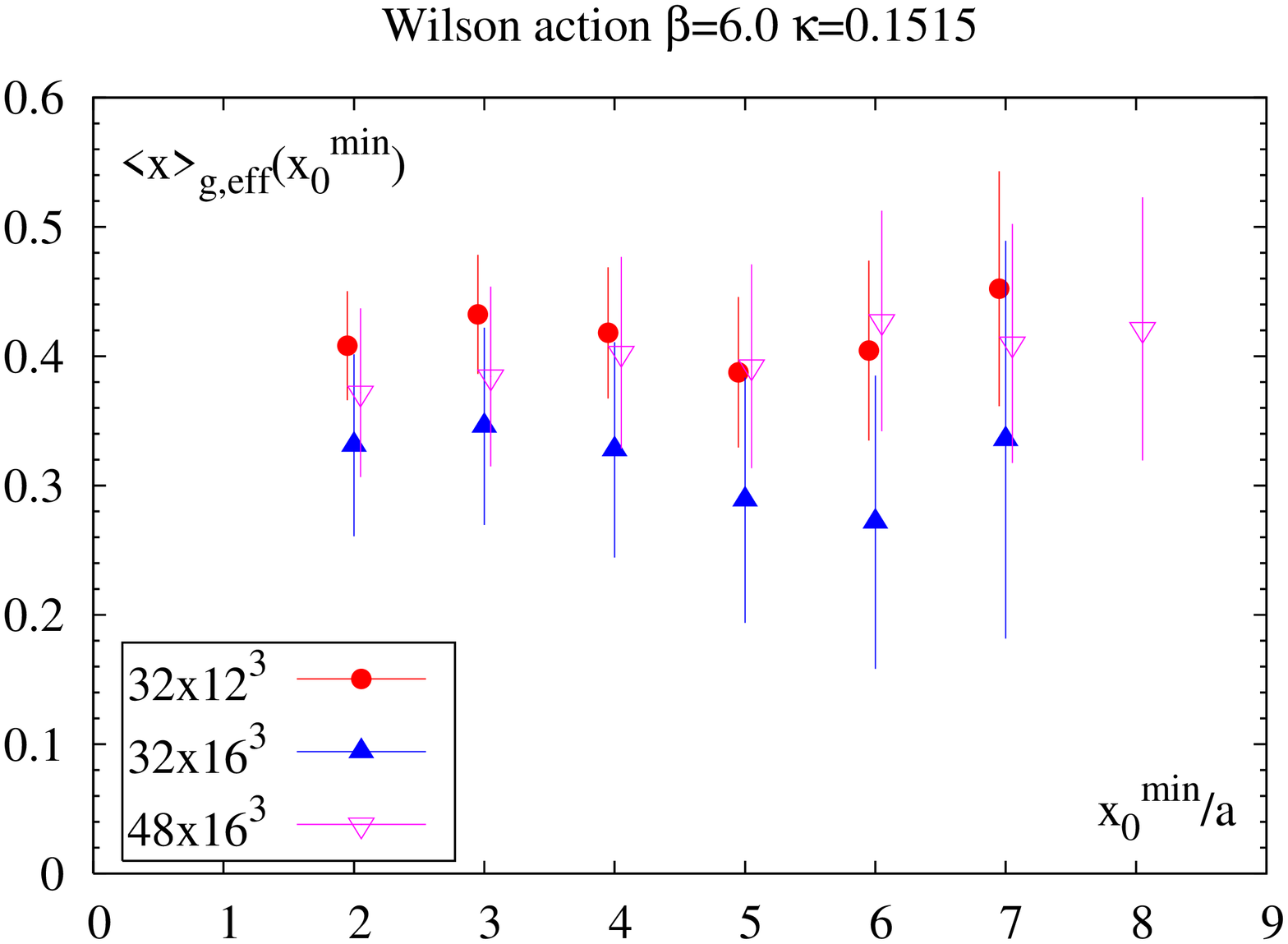,angle=-90,width=8cm}
\end{minipage}
\end{center}
\vspace{-0.5cm}
\caption{\underline{Left}: the relative variance, $\<{\cal O}^2\>/\<{\cal O}\>^2-1$,
of the operator ${\cal O}=\sum_x  \ovl T^{\rm g}_{00}(x)$ as listed in Table 1.
The lines of the form $y=ax^4$ are drawn to guide the eye.
\underline{Right}: The effective gluonic momentum fraction $\<x\>^{(\pi)}_{\rm g,eff}(x^{\rm min}_0)$
  in a heavy pion,  $M_\pi \simeq 1060$MeV, using the HYP-plaquette discretization.}
\la{fig:variance}
\end{figure}

The $\chi$ normalization factors are given in \tab\ref{tab:varT00} for the four discretizations.
We have thus determined the non-perturbative normalization of $\ovl T_{00}^{\rm g}$
in a range of lattice spacings~\cite{necco-sommer}, \mbox{$5<r_0/a<10$.}
As a  check of the correct normalization of the chosen HYP-plaquette
operator, we computed its expectation value on the lightest scalar glueball.
In that case, we know that the momentum fraction carried by the glue is one,
and indeed we find $\<x\>^{(G)}_{\rm g} =1.16(18)$.

\section{The glue momentum fraction in the pion}

We consider a triplet of Wilson quarks, labeled $u,d,s$, with periodic
boundary conditions in all directions and with common $\kappa=0.1515,~0.1530$
and $0.1550$ corresponding to pion masses approximately 1060,
890  and 620 MeV on the $\beta=6$ lattices $32\cdot12^3$, $32\cdot16^3$, $48\cdot16^3$  and $24^4$.
We use $r_0=0.5$fm to set the scale.
To calculate the gluonic momentum fraction in the pion,
we define the effective momentum fraction
\ba\la{eq:xeff}
\<x\>^{(\pi)}_{\rm g,eff}(x^{\rm min}_0) \equiv
\frac{8}{3M_\pi}~ \frac{a^3}{|\Lambda_0|}~ \times  
\sum_{{\bf x};\,x_0\in\Lambda_0}
\left[\frac{   \sum_{\bf y} \<j(0)~ \overline T^{\rm hp}_{00}(x_\odot)~j(\frac{L_0}{2},{\bf y})\>}
     {\sum_{\bf y'} \<j(0)~j(\frac{L_0}{2},{\bf y}')\> }
- \<\overline T^{\rm hp}_{00}(x_\odot) \>
\right],
\ea
where $\Lambda_0 \!= \!\{x_0^{\rm min},\dots,\frac{L_0}{2} \! -\! x_0^{\rm min}\! -\! a,
 \frac{L_0}{2}+x_0^{\rm min},\dots,L_0-x_0^{\rm min}\! -\! a\}$.
For large $L_0$ and $ x_0^{\rm min}$, \mbox{$\<x\>^{(\pi)}_{\rm g,eff} \to \<x\>^{(\pi)}_{\rm g}$.}
As a source field for the pion, we use the isovector pseudoscalar density
$j(x)=\bar d(x)\gamma_5u(x)$. Its two-point function is positive
on every configuration, for each of which we do 12 inversions
corresponding to Dirac and color indices.
On a $24^4$ lattice, we take advantage of the symmetry between all directions
to perform these inversions at the points $k(6,6,6,6)$ for $k=0,1,2,3$
and symmetrize expression (\ref{eq:xeff}) with respect to all directions, so that
$\sum_{x,\mu}\overline T_{\mu\mu}(x)$ vanishes on every  configuration. The right panel of 
Fig.~\ref{fig:variance} shows
our stable plateaus for $\<x\>^{(\pi)}_{\rm g,eff}$ at large values of $ x_0^{\rm min}$ for
three lattice sizes, and the results are summarized in \tab\ref{tab:xg}.

\begin{table}
\begin{center}
\begin{tabular}{|c|c@{~~~}c@{~~~}c@{~~~}c|}
\hline
  $M_\pi$ (Mev)      &  $32\cdot12^3$  & $32\cdot16^3$& $48\cdot16^3$ & $24^4$  \\
\hline
1060(10)   & $0.39(6)_{23091}$   & $0.29(9)_{7113}$ & $0.40(8)_{8331}$  & $0.34(9)_{1048}$ \\
891(9)   &  ---      &  ---   &  ---      & $0.36(8)_{3066}$  \\
624(6)  & ---       &  ---   &  ---      & $0.58(16)_{2538}$   \\
 \hline
\end{tabular}
\end{center}
\vspace{-0.4cm}
\caption{The glue momentum fraction $\<x\>_{\rm g}^{(\pi)}$  in the pion. 
The integer in each subscript denotes the number of configurations used.}
\la{tab:xg}
\end{table}

The lattice equivalent of \eq\ref{eq:xb} has been derived 
for Wilson lattice QCD  in~\cite{hm-sumrules}. It reads
\be
1+{\rm O}(a^2) = \<x\>_{\rm g} +  \<x\>_{\rm f},
\qquad 
 \<x\>_{\rm f} = Z_{\rm f}(g_0) \<x\>^{\rm bare}_{\rm f},~~
 \<x\>_{\rm g} = Z_{\rm g}(g_0) \<x\>^{\rm bare}_{\rm g}.
\la{eq:bare sr}
\ee
The particular discretization of $\ovl T^{\rm g}_{00}$ 
appearing in the lattice sum rule is the bare plaquette one.
The normalization factor $Z_{\rm g}(g_0)$ and its fermionic analog $Z_{\rm f}(g_0)$
are given by the so-called anisotropy coefficients used in thermodynamic studies.
The MSR thus holds even in an `unrenormalized form'
where $\<x\>_{\rm f,g}$  do not separately have a continuum limit;
their renormalization is performed in the next section.

\section{Renormalization of $\<x\>_{\rm g}$ \la{sec:renorm}}

Recall that, in QCD,
the renormalization pattern in the singlet sector reads~\cite{ji-prd}
\be
\left[\begin{array}{c} \ovl T_{00}^{\rm g}(\mu) \\ \ovl T_{00}^{\rm f}(\mu)  \end{array}\right]
= \Bigg[\begin{array}{l@{~~}r} Z_{\rm gg} & 1 \!-\! Z_{\rm ff} \\
                                1\! -\! Z_{\rm gg} & Z_{\rm ff} \end{array}\Bigg]
\left[\begin{array}{c} \ovl T_{00}^{\rm g}(g_0) \\ \ovl T_{00}^{\rm f}(g_0) \end{array}\right],
\ee
provided $\ovl T_{00}^{\rm f,g}(g_0)$ are normalized so that
Eqs.~(\ref{eq:x}--\ref{eq:xb}) hold.
In lattice regularization, this requires the scheme-independent
$Z_{\rm g}(g_0)$ and $Z_{\rm f}(g_0)$ factors, whereas $Z_{\rm gg}$
and $Z_{\rm ff}$ are scheme-dependent functions of $(a\mu,g_0)$.
The renormalization group equation then takes the form
\be
\mu\partial_\mu \!
\left[\begin{array}{c}\! \<x\>_{\rm g}(\mu^2)\! \\ \! \<x\>_{\rm f}(\mu^2)\!  \end{array}\right]
\! =-\bar g^2(\mu)\! \left[\begin{array}{l@{~~}r} c_{\rm gg}(\bar g) & - c_{\rm ff}(\bar g) \\
- c_{\rm gg}(\bar g) & c_{\rm ff}(\bar g) \end{array}\right] \!\!
\left[\begin{array}{c}\! \<x\>_{\rm g}(\mu^2)\! \\ \! \<x\>_{\rm f}(\mu^2)\!  \end{array}\right]
\nonumber
\ee
with $ \mu\partial_\mu \log[Z_{\rm gg} + Z_{\rm ff}-1] = -\bar g^2[ c_{\rm gg}+ c_{\rm ff}]$
and $c_{\rm gg,ff}(\bar g=0)= \frac{N_{\rm f}}{12\pi^2}, ~ \frac{4}{9\pi^2}$
respectively~\cite{gross-wilczek}.
Besides the zero-mode $\ovl T_{00}$, the linear combination
$[1+\tau(\mu)]\ovl T^{\rm g}_{00}(\mu)+\tau(\mu) \ovl T^{\rm f}_{00}(\mu)$
renormalizes multiplicatively with anomalous dimension $-\bar g^2[c_{\rm ff}+ c_{\rm gg}]$,
where $\mu\partial_\mu\tau = -\bar g^2[(c_{\rm ff}+ c_{\rm gg}) \tau + c_{\rm ff}]$.
Note that the asymptotic glue momentum fraction
is given by $c_{\rm ff}(0)/[c_{\rm ff}(0)+c_{\rm gg}(0)]
= Z_{\rm gg}(\infty)=1\! -\! Z_{\rm ff}(\infty)= -\tau(\infty)=16/[16+3N_{\rm f}]$.

In the quenched approximation, $Z_{\rm gg}=1$ due to the absence
of quark loops~\cite{singlet MSbar,singlet SF}. This implies that the singlet quark operator
renormalizes multiplicatively and with the same $Z$-factor, 
computed non-perturbatively in~\cite{continuous external momenta}, as the non-singlet operator:
\ba
\<x\>_{\rm g}(\mu^2) &=& \<x\>_{\rm g} + [1-Z_{\rm ff}(a\mu,g_0)] ~\<x\>_{\rm f},\\
\<x\>_{\rm f}(\mu^2) &=& Z_{\rm ff}(a\mu,g_0) ~ \<x\>_{\rm f}.
\ea

Disregarding disconnected diagrams,
$\<x\>^{\rm bare}_{\rm f}$ has been computed in~\cite{pion matrix element} at the
bare parameters $(\beta=6,\kappa=0.1530)$.
The product $Z_{\rm ff}(a\mu,g_0)Z_{\rm f}(g_0)=0.99(4)$ (in the $\ovl{MS}$-scheme
at $\mu=2$GeV) is known from~\cite{continuous external momenta,pion matrix element}. 
The factor $Z_{\rm f}(g_0)=1+{\rm O}(g_0^2)$ is as yet unknown beyond tree level. 
If we allow for a conservative error,
based on the typical size of one-loop corrections, $Z_{\rm f}(g_0)=1.0(2)$,
our final result is
\be
\<x\>^{(\pi)}_{\rm g}(\mu^2_{\ovl{MS}}=4{\rm GeV}^2) = 0.37(8)(12) \qquad (M_\pi=890{\rm MeV}),
\nonumber
\ee
where the first error is statistical and the second comes from the uncertainty in
$Z_{\rm f}(g_0)$. We observe no significant  quark-mass dependence, 
and given the uncertainties, we do not attempt a chiral extrapolation.

Our result for the glue momentum fraction in a (heavy) pion is compatible with
phenomenological determinations~\cite{SMRS92,gluck},
 $\<x\>^{\ovl{MS}}_g=0.38(5)$ at $Q^2=4{\rm GeV}^2$, based on Drell-Yan,
prompt photoproduction, and the model assumption
that sea quarks carry 10-20\% of the momentum.
The agreement suggests a mild quark-mass dependence, but only a calculation in
full QCD and at smaller masses can substantiate this. Our result at $Q^2=4{\rm GeV}^2$ lies
clearly below the $N_{\rm f}=3$ asymptotic glue momentum fraction of 0.64.
The fact that our result and the valence quark momentum fraction,
computed in~\cite{pion matrix element}, add up to $0.99(8)(12)$
suggests that the omitted disconnected diagrams are small.

\section{Conclusion}
We have computed the glue momentum fraction $\<x\>_{\rm g}$
in a pion of mass $0.6{\rm GeV}<M_\pi<1.06{\rm GeV}$ using quenched lattice QCD
simulations. We find $37(8)(12)\%$ at $\mu_{\ovl{MS}}=2$GeV,
a result compatible with phenomenological determinations~\cite{SMRS92,gluck}.

Although it appears difficult to achieve
precision at the percent level, the present method is applicable to full QCD with dynamical quarks.
Presently the larger uncertainty comes from the
normalization of the quark contribution to the renormalized $\<x\>_{\rm g}$,
and could be reduced significantly by a one-loop calculation.

We also evaluated the bare trace anomaly contribution to the pion's mass
in the same framework~\cite{gluex}.
The counterterm remains to be calculated, but it will ultimately be preferable
to use chiral fermions to avoid mixing with the lower dimensional fermion operator.

Finally, we remark that
the freedom of choosing a numerically advantageous discretization
of $\ovl T_{\mu\nu}$ has not been fully exploited
in previous lattice simulations.
The improvement that was essential in the present computation of the pion
momentum fraction can be carried over
to fully dynamical calculations and the exploration of other observables,
such as the gluon contribution to the nucleon spin.   It is also particularly
promising for thermodynamic studies of pressure,
energy density and transport coefficients.

\noindent\textit{Acknowledgments.~~}
This work was supported in part by
funds provided by the U.S. Department of Energy under cooperative research agreement
DE-FG02-94ER40818.


\end{document}